\documentclass[11pt]{article}
\newcommand{\ssc}{\scriptscriptstyle}
\newcommand{\be}{\begin{equation}}
\newcommand{\ee}{\end{equation}}
\newcommand{\bea}{\begin{eqnarray}}
\newcommand{\eea}{\end{eqnarray}}
\usepackage{amscd}
\usepackage{amssymb}
\usepackage{amsfonts}
\usepackage{amsmath}
\usepackage{mathrsfs}
\begin{document}

\setcounter{page}{1}
\vfil

\pagestyle{plain}

\begin{center}
{\LARGE {\bf Quantization as a dimensional reduction phenomenon}}

\bigskip
\bigskip
\bigskip

{\large E. Gozzi} and {\large D. Mauro}

             Department of Theoretical Physics, 
	     University of Trieste, \\
	     Strada Costiera 11, Miramare-Grignano, 34014 Trieste, Italy,\\
	     and INFN, Trieste, Italy\\
	     e-mail: {\it gozzi@ts.infn.it},  {\it mauro@ts.infn.it}

\end{center}

\bigskip
\bigskip
\begin{abstract}
Classical mechanics, in the operatorial formulation of Koopman and von Neumann, can be written also in a functional form. In this form two Grassmann partners of time make their natural appearance extending in this manner time to a three dimensional supermanifold. Quantization is then achieved by a process of dimensional reduction of this supermanifold. We prove that this procedure is equivalent to the well-known method of geometric quantization.\footnote{Talk given by E.G. at the conference ``{\it Are there Quantum Jumps? - On the Present Status of Quantum Mechanics}'' in honour of the 70th birthday of Giancarlo Ghirardi.}
\end{abstract}


\section{Prelude}

It is a pleasure to dedicate this work to Giancarlo Ghirardi on the occasion of his 70th birthday. We have never collaborated with Giancarlo but he has been for us an example of a physicist who has never given up thinking about quantum mechanics (QM). He did so even when ``thinking about QM'' was considered unfashionable. 

The geometric approach to QM that we will present here is not close to anything Giancarlo did, nevertheless he has always been very supportive of our work. We want to warmly thank him for that support and wish him many happy returns. 

\section{Introduction}

Quantum mechanics is for sure the most counterintuitive theory physicists have ever invented. At the same time it is the best tested experimentally. Also relativity (both special and general) is a counterintuitive theory with its ``strange'' phenomena of length contraction, time dilation and so on. Anyhow physicists got accustomed to these relativity weird phenomena because a geometric framework was provided immediately after their introduction: it was Minkowski geometry for special relativity and Riemann geometry for general relativity. Even if these ``non standard'' geometries were different from the Euclidean one which we humans were accustomed to, as slow-moving objects, still a ``non-standard'' geometric interpretation of relativity helped in accepting those strange phenomena. It helped because geometries, even the non Euclidean ones, are something humans can ``somehow'' visualize and get accustomed to. Nothing similar has happened for QM. There the only tools used are basically operators and Hilbert spaces, which do not have any intuitive interpretation in terms of the {\it geometry} of spacetime. 
The first time QM was brought close to space-time was via the path integral formulation of R.P. Feynman \cite{uno} whose article carried, significantly, the title ``{\it A space-time approach to QM}''. A better {\it geometrical} understanding of QM would be useful not only in order to get a deeper grasp of its counterintuitive phenomena but also for another important reason. We know in fact that gravity, which is the queen of ``geometrical theories'', develops problems when one tries to marry it with QM (i.e. to quantize it). This to us seems to suggest that the ``geometry'' hidden behind QM is very different than the standard Riemann geometry of general relativity. We feel this {\it hidden weird geometry} may be responsible for many of the counterintuitive phenomena of QM. 

As the first approach to QM from a geometric or space-time point of view has been the one of Feynman, we decided to continue along the same lines. In this paper we will present a path-integral approach \cite{due} even for classical mechanics (CM).
This can be done once CM is formulated in an operatorial form like 
Koopman and von Neumann \cite{tre} did in the Thirties. From now on we will indicate this
path integral formulation of {\it classical} mechanics with the acronym ``{\it CPI}'' 
({\it C}lassical {\it P}ath {\it I}ntegral) while we will use for the Feynman {\it P}ath {\it I}ntegral of {\it Q}uantum mechanics the symbol ``{\it QPI}''. The operatorial approach to CM developed by {\it K}oopman and {\it v}on {\it N}eumann will be instead indicated with the acronym {\it KvN}.

The CPI can be put in a very compact form by extending time to a 
3-dimensional supertime made of $t$ and of two 
Grassmann partners $\theta, \bar{\theta}$. Quantization, which means a manner to go from the CPI to the QPI, is achieved by properly freezing to zero the two Grassmann partners
of time $(\theta, \bar{\theta})$. This procedure of quantization via a contraction from 3-dim to 1-dim, resembles very much the mechanisms used in Kaluza-Klein and string 
theories where one starts from an 11-dim theory and brings it down, via dimensional reduction,
to 4-dim. It is amazing that quantization itself can be obtained via a dimensional reduction procedure. Of course, it is a dimensional reduction that involves Grassmann variables and so all the geometry
that one gets is very hard to visualize. This ``hidden, hard-to-visualize {\it geometry}''
may be the structure that, as we said at the beginning of this section, is at the origin of all
the ``strange'' phenomena of QM. 

More direct is the procedure of ``{\it dequantization}'', which means a manner to go from the QPI to the CPI. This can be achieved via an extension of time $t$
to supertime $(t,\theta,\bar{\theta})$ and of phase space to a superphase space, which appears
naturally \cite{due} in the path integral approach to CM. We call ``{\it geometrical}'' this procedure
of dequantization because it involves geometrical structures like time and its Grassmann
partners. At the relativistic field theory level we expect that the role played by time will 
be taken by space-time and this is another reason for the use of the word ``{\it geometrical}'' in our procedure.

\section{Brief Review of the KvN Formalism and of the CPI}

In the Thirties Koopman and von Neumann \cite{tre}, triggered most probably 
by the operatorial aspects of QM, proposed an operatorial formulation
for CM. Their work is basically an extension of the one of Liouville 
who, for classical statistical mechanics, had found the equation of evolution 
for the probability distribution $\rho(p,q)$ defined on phase space. 
The Liouville equation is:
\begin{equation}
\displaystyle i\frac{\partial}{\partial t}\rho =\hat{L}\rho \label{8-1}
\end{equation}
where $\hat{L}$ is the Liouvillian defined as 
\begin{displaymath}
\displaystyle \hat{L}=-i\partial_pH\partial_q+i\partial_qH\partial_p 
\end{displaymath}
with $H$ the Hamiltonian function of the system. We will use
the following notation: ${\cal M}$ indicates the $2n$ dimensional phase space, while $\varphi^a
=(q^1 \cdots q^n,p^1 \cdots p^n)$ with $a=1,\cdots, 2n$ indicate the $2n$ phase
space coordinates. Liouville's probability distributions $\rho(\varphi)$  
are $L^1$ functions because the only request made on them is to be integrable
\begin{displaymath}
\displaystyle \int d \varphi \; \rho(\varphi)=1
\end{displaymath}
where by ``$d\varphi$'' we mean ``$d^{2n}\varphi$''.
Koopman and von Neumann, instead of using the probability distribution
$\rho(\varphi)$, introduced a {\it Hilbert space} of $L^2$ functions
$\psi(\varphi)$. Furthermore they imposed the following two postulates for $\psi(\varphi)$:

{\bf 1st postulate}: The equation of evolution for $\psi$ is
\begin{equation}
i\frac{\partial\psi}{\partial t}=\hat{L}\psi, \label{9-2}
\end{equation}
that is the same as the Liouville equation (\ref{8-1}). 

{\bf 2nd postulate}: The probability distribution $\rho(\varphi)$ can be built out of the $\psi$ as follows:
\begin{equation}
\rho(\varphi)=|\psi(\varphi)|^2. \label{10-1}
\end{equation}
These two postulates are not in contradiction with the Liouville equation
(\ref{8-1}). In fact, since $\hat{L}$ is a differential operator, linear in 
the first order derivatives, from (\ref{9-2}) and (\ref{10-1}) it is easy to
derive (\ref{8-1}), that means that the same equation is satisfied by both
$\rho$ and $\psi$. This would not happen in QM, where the KvN equation
(\ref{9-2}) is replaced by the Schr\"odinger one and the quantum equation
for $\rho$ is a continuity equation, different from the Schr\"odinger one. This is due to
the presence of second order derivatives in the Schr\"odinger operator.

In \cite{sette} we called the $\psi$ of (\ref{9-2}) ``{\it the KvN waves}'' and
we thoroughly analyzed their physics. The overall KvN formalism 
is formally very similar to the one
of QM, even if the Liouvillian $\hat{L}$ \cite{sette}
reproduces the {\it classical} dynamics. Like every operatorial formalism also
the KvN one has a path integral counterpart, which has been fully developed in Ref.
\cite{due} and it reproduces the
kernel of evolution associated to the classical equation (\ref{9-2}). In CM if we ask which is the transition amplitude $K(\varphi^a,t|
\varphi^a_{\ssc 0},t_{\ssc 0})$ of arriving in $\varphi^a$ at time $t$ having started from $\varphi^a_{\ssc 0}$ at time $t_{\ssc 0}$, the answer is 
\begin{equation}
\displaystyle K(\varphi^a,t|\varphi^a_{\ssc 0},
t_{\ssc 0})=\delta[\varphi^a-\phi^a_{\textrm{cl}}(t;\varphi_{\ssc 0},t_{\ssc 0})]  \label{11-1}
\end{equation}
where $\phi^a_{\textrm{cl}}(t;\varphi_{\ssc 0},t_{\ssc 0})$ is the solution of Hamilton's equations with initial conditions $\varphi_{\ssc 0}$ 
\begin{equation}
\displaystyle \dot{\varphi}^a=\omega^{ab}\frac{\partial H}{\partial \varphi^b}
\label{12-1} 
\end{equation}
and $\omega^{ab}$ indicates the symplectic matrix \cite{otto}.
We can rewrite (\ref{11-1}) as a sum over all possible intermediate 
configurations, after having sliced the interval $(t-t_{\ssc 0})$ in $N$ intervals
\begin{eqnarray}
K(\varphi^a,t|\varphi_{\ssc 0}^a,t_{\ssc 0})\hspace{-0.4cm}&=& \hspace{-0.4cm} \sum_{\varphi_i}K(\varphi^a,t|\varphi_{\ssc N-1},t_{\ssc N-1})
K(\varphi_{\ssc N-1},t_{\ssc N-1}|\varphi_{\ssc N-2},t_{\ssc N-2})\cdots K(\varphi_{\ssc 1},t_{\ssc 1}|\varphi^a_{\ssc 0},t_{\ssc 0})\nonumber\\
\hspace{-0.4cm}&=&\hspace{-0.4cm} \prod_{j=1}^N\int d\varphi_j \,\delta[\varphi^a_j-\phi^a_{\textrm{cl}}
(t_j;\varphi_{j-1},t_{j-1})]  \label{12-2} \\
&\xrightarrow{N\rightarrow\infty}& \int {\mathscr D}\varphi \,
\widetilde{\delta}
[\varphi^a-\phi_{\textrm{cl}}^a(t;\varphi_{\ssc 0},t_{\ssc 0})]. \nonumber
\end{eqnarray}
In the first line of (\ref{12-2}) $|\varphi_i,t_i)$ denotes a set of intermediate configurations
between $|\varphi_{\ssc 0}^a,t_{\ssc 0})$ and $|\varphi^a,t)$, while in the last one 
we have sent the number of slices 
to infinity recovering in this manner a functional integration. 
The $\widetilde{\delta}[\qquad]$ indicates a functional Dirac delta which effectively gives 
weight ``one'' to the classical trajectory and weight ``zero'' to all the other ones. 
This functional delta can be rewritten as a delta on the Hamilton equations 
of motion (\ref{12-1}) via the introduction of a suitable functional determinant 
\begin{equation}
\displaystyle \widetilde{\delta}[\varphi^a-\phi_{\textrm{cl}}^a(t;\varphi_{\ssc 0},t_{\ssc 0})]=
\widetilde{\delta} [\dot{\varphi}^a-\omega^{ab}\partial_bH]\textrm{det}(
\delta_b^a\partial_t-\omega^{ac}\partial_c\partial_bH). \label{13-1}
\end{equation}
Let us now perform a Fourier transform of the Dirac delta on the RHS of (\ref{13-1}) 
introducing $2n$ extra variables $\lambda_a$ and let us exponentiate the determinant
using $4n$ Grassmann variables $c^a$, $\bar{c}_a$. The final result is the
following one:
\begin{equation}
\displaystyle K(\varphi^a,t|\varphi^a_{\ssc 0},t_{\ssc 0})=\int {\mathscr D}^{\prime\prime}
\varphi {\mathscr D}\lambda {\mathscr D} c {\mathscr D}\bar{c}
\; \textrm{exp} \left[i\int_{t_{\ssc 0}}^{t} d\tau {\cal L}\right]
\label{13-2}
\end{equation} 
where ${\mathscr D}^{\prime\prime}\varphi$ indicates that the integration 
in $\varphi$ is over the paths $\varphi(t)$ with fixed end points $\varphi_{\ssc 0}$ and
$\varphi$, while the other integrations ${\mathscr D}\lambda {\mathscr D}c{\mathscr
D}\bar{c}$ include the end points of the paths in $\lambda$, $c$, $\bar{c}$.
The ${\cal L}$ which appears in (\ref{13-2}) is
\begin{equation}
\displaystyle {\cal L}=\lambda_a\dot{\varphi}^a+i\bar{c}_a\dot{c}^a
-\lambda_a\omega^{ab}\partial_bH-i\bar{c}_a\omega^{ad}\partial_d\partial_bHc^b
\label{14-1a}
\end{equation}
and its associated Hamiltonian is:
\begin{equation}
{\cal H}=\lambda_a\omega^{ab}\partial_bH+i\bar{c}_a\omega^{ad}
\partial_d\partial_bHc^b \label{14-1b}.
\end{equation}
The path integral (\ref{13-2}) has an operatorial counterpart and we shall now check
that this is the KvN theory. Let us first remember \cite{uno} that from the kinetic
part of the Lagrangian appearing in the weight of a path integral, one can deduce the 
commutators of the associated operator theory. In our case from (\ref{14-1a}) we get the following graded commutators:
\begin{equation}
[\hat{\varphi}^a,\hat{\lambda}_b]=i\delta_b^a, \qquad \qquad
[\hat{c}^a,\hat{\bar{c}}_b]=\delta_b^a \label{14-2}
\end{equation}
while all the others are zero, in particular $[\hat{p},\hat{q}]=0$ and this is the
clear indication that what we have obtained is CM and not QM. Despite having objects that do not commute, as indicated by (\ref{14-2}), in the CPI there are no ordering problems. This issue is analyzed in detail in Ref. \cite{ann}.

The commutators (\ref{14-2}) can be realized in various ways. One way is to implement
$\hat{\varphi}^a$ and $\hat{c}^a$ as multiplicative operators and $\hat{\lambda}_a$ and $\hat{\bar{c}}_b$ as derivative ones:
\begin{equation}
\displaystyle \hat{\lambda}_a=-i\frac{\partial}{\partial\varphi^a}, \qquad
\hat{\bar{c}}_b=\frac{\partial}{\partial c^b} \label{14-3}.
\end{equation}
This realization automatically defines a basis $|\varphi^a,c^a\rangle$ given by:
\begin{equation}
\hat{\varphi}^a|\varphi^a,c^a\rangle =\varphi^a|\varphi^a,c^a\rangle, \qquad 
\hat{c}^a|\varphi^a,c^a\rangle=c^a|\varphi^a,c^a\rangle. \label{14-4}
\end{equation}
The realization (\ref{14-3}) is not the only possible one. Among the $4n$ operators 
$\hat{\varphi}^a\equiv (\hat{q},\hat{p})$ and 
$\hat{\lambda}_a\equiv (\hat{\lambda}_q,\hat{\lambda}_p)$ we could have chosen
to implement $\hat{q}$ and $\hat{\lambda}_p$ as multiplicative operators
and, as they commute because of the commutators (\ref{14-2}), diagonalize them simultaneously. As a consequence $\hat{p}$ and $\hat{\lambda}_q$ must then be realized as derivative operators:
\begin{displaymath}
\displaystyle \hat{p}=i\frac{\partial}{\partial \lambda_p}, \qquad \quad
\displaystyle \hat{\lambda}_q=-i\frac{\partial}{\partial q}. \label{15-1}
\end{displaymath}
The same can be done for $\hat{c}^a\equiv (\hat{c}^q,\hat{c}^p)$ and $\hat{\bar{c}}_a
\equiv (\hat{\bar{c}}_q,\hat{\bar{c}}_p)$. The basis defined by this overall
realization is 
\begin{equation}
\begin{array}{l}
\hat{q}|q,\lambda_p,c^q,\bar{c}_p\rangle
=q|q,\lambda_p,c^q,\bar{c}_p\rangle \medskip\\
\hat{\lambda}_p|q,\lambda_p,c^q,\bar{c}_p\rangle=
\lambda_p|q,\lambda_p,c^q,\bar{c}_p\rangle \medskip\\
\hat{c}^q|q,\lambda_p,c^q,\bar{c}_p\rangle=
c^q|q,\lambda_p,c^q,\bar{c}_p\rangle \medskip \\
\hat{\bar{c}}_p|q,\lambda_p,c^q,\bar{c}_p\rangle=
\bar{c}_p|q,\lambda_p,c^q,\bar{c}_p\rangle. \label{15-2}
\end{array}
\end{equation}
Let us for the moment stick to the realization (\ref{14-3})-(\ref{14-4}) and build
the {\it operator} associated with the Hamiltonian (\ref{14-1b}). If we restrict
ourselves to its non-Grassmann part, which
we will indicate with ${\cal H}_{\ssc B}$ (B for Bosonic), we get:
\begin{displaymath}
\displaystyle {\cal H}_{\ssc B}=\lambda_a\omega^{ab}\partial_bH\, \longrightarrow
\, \hat{{\cal H}}_{\ssc B}=-i\omega^{ab}\partial_bH\partial_a=-i\partial_p
H\partial_q+i\partial_qH\partial_p=\hat{L}.
\end{displaymath}
This tells us that $\hat{{\cal H}}_{\scriptscriptstyle B}$ is exactly the Liouville operator, so the kernel (\ref{13-2}), in its bosonic part, can be written as:
\begin{equation}
\displaystyle K(\varphi, t|\varphi_{\ssc 0},t_{\ssc 0})=\langle \varphi|\exp
-i\hat{L}(t-t_{\ssc 0})|\varphi_{\ssc 0} \rangle \label{16-2}.
\end{equation} 
This confirms (modulo the Grassmann part) that the path integral
(\ref{13-2}) is exactly the functional counterpart of the KvN theory.

What can we say about the Grassmann part of ${\cal H}$? This has been thoroughly 
analyzed in Refs. \cite{due} and \cite{quattro}. Here we will be rather brief,
referring the reader to papers \cite{due} and \cite{quattro} for further details. Once
the operatorial realization (\ref{14-3}) is used, the Hamiltonian ${\cal H}$
in (\ref{14-1b}) is turned into the following operator:
\begin{equation}
\displaystyle \hat{{\cal H}}=-i\omega^{ab}\partial_bH\partial_a
-i\omega^{ab}\partial_b\partial_dH\hat{c}^d\frac{\partial}{\partial c^a}.
\label{16-3}
\end{equation}
Via the commutators (\ref{14-2}) the equations of motion of the $\hat{c}^a$ turn out
to be
\begin{displaymath}
\dot{\hat{c}}^a=i[{\cal H}, \hat{c}^a]=\omega^{ab}\partial_b\partial_dH\hat{c}^d. \label{17-1}
\end{displaymath}
Note that these are the same equations of motion of the differentials $d\hat{\varphi}^a$,
which can be obtained by doing the first variation of the Hamilton equations
(\ref{12-1}). So we can make the identification 
\be
\displaystyle \hat{c}^a=d\hat{\varphi}^a, \label{17-2}
\ee
that means the $\hat{c}^a$ are essentially a basis for the differential forms of the phase space \cite{otto}. In general, a form has the expression 
\be
\displaystyle F(\hat{\varphi}, d\hat{\varphi})=
F_{\ssc 0}(\hat{\varphi})+F_a(\hat{\varphi})d\hat{\varphi}^a+F_{ab}(\hat{\varphi})d
\hat{\varphi}^a\wedge d\hat{\varphi}^b+\cdots, \label{17-3}
\ee
where ``$\wedge$'' is the antisymmetric wedge product usually defined among forms
\cite{otto}. Via the identification (\ref{17-2}) the form (\ref{17-3}) becomes a 
function of $\hat{c}$ 
\be
\displaystyle F(\hat{\varphi}, d\hat{\varphi})\,\longrightarrow \,
F(\hat{\varphi}, \hat{c})=F_{\ssc 0}(\hat{\varphi})+F_a(\hat{\varphi})\hat{c}^a+F_{ab}(\hat{\varphi})
\hat{c}^a\hat{c}^b+\cdots \label{17-4}
\ee
where there is no need to introduce the wedge product anymore, because this is
automatically taken care of by the Grassmann character of the operators $\hat{c}^a$. Like $\hat{c}^a$ can be given a geometrical interpretation, the same can be done for the 
$\hat{{\cal H}}$ in (\ref{16-3}). This turns out \cite{due}, \cite{quattro}
to be nothing else than the Lie derivative along the Hamiltonian vector field 
\cite{otto} associated with $H$. So this is basically the geometrical object which extends
the Liouville operator to the space of forms. While the Liouville operator
makes the evolution only of the zero forms that are the $F_{\ssc 0}(\hat{\varphi})$ of (\ref{17-4}), the Lie derivative $\hat{{\cal H}}$ makes the evolution of
the entire $F(\hat{\varphi},\hat{c})$ with all its components appearing on the 
RHS of (\ref{17-4}). Besides the forms and the Lie derivatives, also many other
geometrical structures, like Lie brackets, exterior derivatives, etc. \cite{otto}
can be written in the language of our CPI and this has been done in great details 
in Refs. \cite{due} and \cite{quattro}. 

Having now these extra variables $c^a$, the ``Koopman-von Neumann waves''
$\psi(\varphi)$ will be extended to functions of both $\varphi$ and $c$: $\psi(\varphi,c)$.
Of course, in order to have a Hilbert space, a proper scalar product 
among these extended waves must be given. All this has been done in detail in Ref. \cite{sette}.
Using the basis defined in (\ref{14-4}) and a proper scalar product we could represent
the ``KvN waves'' $\psi(\varphi,c)$ as:
\be
\displaystyle \langle \varphi^a,c^a|\psi\rangle =\psi(\varphi^a,c^a). \label{19-1}
\ee
In this basis we could also generalize the kernel of propagation (\ref{16-2}) 
to the following one 
\be
\displaystyle K(\varphi,c,t|\varphi_{\ssc 0},c_{\ssc 0},t_{\ssc 0})
\equiv \langle \varphi,c,t|\varphi_{\ssc 0},c_{\ssc 0},t_{\ssc 0}\rangle=\langle \varphi,c|
\exp -i\hat{{\cal H}}(t-t_{\ssc 0})|\varphi_{\ssc 0},c_{\ssc 0}\rangle, \label{19-2}
\ee
whose path integral representation is
\be
\displaystyle K(\varphi,c,t|\varphi_{\ssc 0},c_{\ssc 0},t_{\ssc 0})
=\int {\mathscr D}^{\prime\prime} \varphi {\mathscr D}\lambda
{\mathscr D}^{\prime\prime}c{\mathscr D}\bar{c} \, \exp i\int_{t_{\ssc 0}}^t 
d\tau {\cal L} \label{19-3}.
\ee
The difference with respect to (\ref{13-2}) is in the measure of integration,
which in (\ref{19-3}) has the initial and final $c$ {\it not} integrated over.

\section{Supertime and Superphase space}

We have seen in the previous chapter how the $8n$ variables $(\varphi^a,\lambda_a,c^a,\bar{c}_a)$,
which enter the Lagrangian (\ref{14-1a}), can be turned into operators by the path integral 
(\ref{13-2}). Actually if one looks at ${\cal L}$ and ${\cal H}$ not as weight
of a path integral but as a standard Lagrangian and Hamiltonian of a classical system, one could then look
at $(\varphi^a,\lambda_a,c^a,\bar{c}_a)$ as coordinates of an {\it extended phase space}.
The variation of the action $\int d\tau {\cal L}$ with respect to $\lambda_a$,
$\bar{c}_a$, $c^a$, $\varphi^a$ would give respectively the following equations of motion: 
\begin{equation}
\begin{array}{l}
\dot{\varphi}^a=\omega^{ab}\partial_bH \medskip \\
\dot{c}^a=\omega^{ad}\partial_d\partial_bH c^b \medskip \\
\dot{\bar{c}}_b=-\bar{c}_a\omega^{ad}\partial_d\partial_bH \medskip \\
\dot{\lambda}_b=-\omega^{ad}\partial_d\partial_bH\lambda_a-i\bar{c}_a\omega^{ad}
\partial_d\partial_f\partial_bHc^f. \label{20-1}
\end{array}
\end{equation}
They could be obtained also via the ${\cal H}$ of (\ref{14-1b}) by postulating the following {\it e}xtended {\it P}oisson {\it b}rackets (epb):
\bea
\displaystyle \{\varphi^a,\lambda_b\}_{epb}=\delta_b^a, && \qquad  \{\bar{c}_b,c^a\}_{epb}=-i\delta_b^a, \medskip \label{3.2} \\
&&\displaystyle  \hspace{-1cm} \{\varphi^a,\varphi^b\}_{epb}=0. \label{3.3}
\label{20-2}
\eea 
Note that Eq. (\ref{3.3}) is different from the {\it P}oisson {\it b}rackets (pb)
defined in the {\it non-extended} phase space which are:
\be
\displaystyle \{ \varphi^a,\varphi^b\}_{pb}=\omega^{ab}. \label{3-3bis}
\ee
So we are basically working in an extended phase space. Its precise geometrical structure is the one of a double bundle over the basic phase space ${\cal M}$ coordinatized by the variables $\varphi^a$. This
double bundle has been studied in detail in Ref. \cite{quattro}.

The reader may dislike the pletora of variables $(\varphi^a,\lambda_a,c^a,\bar{c}_a)$
that make up the {\it extended phase space} described above. Actually, thanks to the beautiful geometry 
underlying this space, we can assemble together the $8n$ variables $(\varphi^a,\lambda_a,c^a,\bar{c}_a)$
in a single object as if they were the components of a multiplet. In order to do that we have first to
introduce two Grassmann partners $(\theta,\bar{\theta})$ of the time $t$. The triplet
\be
\displaystyle (t,\theta,\bar{\theta}) \label{23-0}
\ee
is known as {\it supertime} and it is, for the point particle dynamics, the analog of the {\it superspace} 
introduced in supersymmetric field theories \cite{nove}. The object that we mentioned above and which
assembles together the $8n$ variables $(\varphi,\lambda,c, \bar{c})$ is defined as
\be
\displaystyle \Phi^a(t,\theta,\bar{\theta})\equiv \varphi^a(t)+\theta c^a(t)+\bar{\theta}
\omega^{ab}\bar{c}_b(t)+i\bar{\theta}\theta\omega^{ab}\lambda_b(t). \label{21-1}
\ee
We could call the $\Phi^a$ {\it superphase space} variables because their first components $\varphi^a$ 
are the standard phase space variables of the system.
The Grassmann variables $\theta$, $\bar{\theta}$ are complex in the sense in which the operation of complex conjugation can be defined \cite{dieci} for Grassmann variables. Further details can be found in Ref. \cite{ann}, where we also study the dimensions of these Grassmann variables. The main result is that, even if there is a lot of freedom in choosing these dimensions, the combination $\theta\bar{\theta}$ has always the dimensions of an action. Using the superfields the relations (\ref{3.2})-(\ref{3.3}) can be written in a compact form as:
\begin{displaymath}
\{\Phi^a(t,\theta, \bar{\theta}),\Phi^b(t,\theta^{\prime},\bar{\theta}^{\prime})\}_{epb}=
-i \omega^{ab}\delta(\bar{\theta}-\bar{\theta}^{\prime})\delta(\theta-\theta^{\prime}),
\end{displaymath}
which is the extended phase space analog of (\ref{3-3bis}). 

Despite the formal unification of the $8n$ variables $(\varphi,c,\bar{c},\lambda)$
into the $2n$ objects $\Phi^a(t,\theta,\bar{\theta})$ of (\ref{21-1}), the reader may still wonder
why we need $8n$ variables when we know that CM can be described using only the basic phase space variables $\varphi^a$. The answer lies in the fact that we want an object, like ${\cal H}$, which makes at the same time the evolution of the points of the phase space $\varphi$ and of the forms $d\varphi$. But we know that the evolution of the forms $d\varphi$ can actually be derived from the evolution of the points by doing its first variation and this implies that somehow the variables $(\varphi,c,\bar{c},\lambda)$ are redundant. This redundancy is signalled by a set of universal symmetries present in our formalism whose form can be found in Refs. \cite{due}, \cite{quattro}.
The two charges we are most interested in are $\hat{Q}_{\scriptscriptstyle H}$ and 
$\hat{\bar{Q}}_{\scriptscriptstyle H}$ defined as
\begin{equation}
\displaystyle \hat{Q}_{\scriptscriptstyle H}\equiv i\hat{c}^a\hat{\lambda}_a-\hat{c}^a\partial_aH, \qquad 
\displaystyle \hat{\bar{Q}}_{\scriptscriptstyle H}\equiv i\hat{\bar{c}}_a
\omega^{ab}\hat{\lambda}_b+\hat{\bar{c}}_a\omega^{ab}\partial_bH. \label{23-1}
\end{equation}
They make up a ``{\it universal}''
$N=2$ supersymmetry (susy), present for any system. Their anticommutator is:
\begin{displaymath}
\displaystyle [\hat{Q}_{\scriptscriptstyle H},\hat{\bar{Q}}_{\scriptscriptstyle H}]=
2i\hat{\cal H}. 
\end{displaymath}
The charges in (\ref{23-1}) act via graded commutators on the variables $(\hat{\varphi}^a,\hat{c}^a, \hat{\bar{c}}_a,\hat{\lambda}_a)$ which can be considered as the ``{\it target space}'' variables (in modern language), while the ``{\it base space}'' is given by the supertime $(t,\theta,\bar{\theta})$. It is then natural to ask whether the susy operators (\ref{23-1}) are the ``representation'' on the target space of some differential operators $\Omega_{\scriptscriptstyle H}, \bar{\Omega}_{\scriptscriptstyle H}$ acting on the base space. 
The answer \cite{ann} is
\begin{displaymath}
\displaystyle \Omega_{\scriptscriptstyle H}=-\frac{\partial}{\partial \theta}-\bar{\theta}\frac{\partial}{\partial t}, \qquad \bar{\Omega}_{\scriptscriptstyle H}=\frac{\partial}{\partial \bar{\theta}}+\theta \frac{\partial}{\partial t}
\end{displaymath}
and one notices that the susy transformations ``rotate'' $t$ into combinations of $\theta,\bar{\theta}$ and in this sense we called $(\theta, \bar{\theta})$ partners of time. 

A further aspect of $(\theta,\bar{\theta})$
which is worth being explored is the following one. We know that, for an operator theory, we can define either the Heisenberg picture or the Schr\"odinger one. In the KvN version of CM the operators, which will be indicated respectively as $\hat{O}_{\ssc H}(t)$ and $\hat{O}_{\ssc S}$ in the two pictures, are related to each other in the following manner:
\begin{displaymath}
\displaystyle \hat{O}_{\ssc H}(t)\equiv \exp \Bigl[i\hat{{\cal H}}t\Bigr] 
\hat{O}_{\ssc S} \exp \Bigl[-i\hat{{\cal H}}t\Bigr].
\label{27-1}
\end{displaymath}
One question to ask is what happens if we build the Heisenberg picture associated with the partners of time
$\theta,\bar{\theta}$. The analog, for $\theta$ and $\bar{\theta}$, of the usual time translation operator $\hat{{\cal H}}$ is given respectively by $\hat{Q}\equiv i\hat{c}^a\hat{\lambda}_a$ and 
$\hat{\bar{Q}}\equiv i\hat{\bar{c}}_a\omega^{ab}\hat{\lambda}_b$ and so the Heisenberg picture in $\theta$, $\bar{\theta}$ of an operator $\hat{O}_{\ssc S}$ is
\be
\displaystyle \hat{O}_{\ssc H}(\theta,\bar{\theta}) \equiv \exp \left[\theta \hat{Q}+\hat{\bar{Q}}\bar{\theta}\right]
\hat{O}_{\ssc S} \exp \left[-\theta \hat{Q}-\hat{\bar{Q}}\bar{\theta}\right]. \label{connect}
\ee
A simple example to start from is the phase space operator $\hat{\varphi^a}(t)$ which does not depend on $\theta,\bar{\theta}$, so it could be considered as an operator in the Schr\"odinger picture with respect to $\theta,\bar{\theta}$ and in the Heisenberg picture with respect to $t$. Its Heisenberg picture version in $\theta,\bar{\theta}$
can be worked out easily \cite{ann} and the result is:
\begin{displaymath}
\hat{\varphi}^a_{\ssc H}(t)\equiv \exp \Bigl[ \theta\hat{Q}+\hat{\bar{Q}}\bar{\theta}\Bigr] \hat{\varphi}_{\ssc S}^a(t)
\exp \Bigl[ -\theta \hat{Q}-\hat{\bar{Q}}\bar{\theta}\Bigr]
=\hat{\Phi}^a(t,\theta,\bar{\theta}).
\end{displaymath}
This means the {\it superphase space operators} $\hat{\Phi}^a$ {\it can be considered as the Heisenberg picture} version in $\theta$, $\bar{\theta}$ of the phase space operators.
The same holds for any function $G$ of the operators $\hat{\varphi}$, i.e.:
\be
\displaystyle \exp \Bigl[\theta \hat{Q}+\hat{\bar{Q}}\bar{\theta}\Bigr] G(\hat{\varphi}^a)
\exp \Bigl[-\theta\hat{Q}-\hat{\bar{Q}}\bar{\theta}\Bigr] =G(\hat{\Phi}^a). \label{28-1}
\ee
In particular, if the function $G(\hat{\varphi}^a)$ is the Hamiltonian $H(\hat{\varphi}^a)$ we get from (\ref{28-1})
\be
\displaystyle SH(\hat{\varphi}^a)S^{-1} =H[\hat{\Phi}^a] \label{29-1}
\ee
where $S=\exp [\theta \hat{Q} +\hat{\bar{Q}}\bar{\theta}]$. 

At this point it is instructive to expand the RHS of (\ref{29-1}) in terms of $\theta$, $\bar{\theta}$. We get:
\be
\displaystyle H[\hat{\Phi}^a]=H[\hat{\varphi}^a]+\theta \hat{N}+\hat{\bar{N}}\bar{\theta}-
i\bar{\theta}\theta \hat{\cal H} \label{29-2}
\ee
where 
\begin{displaymath}
\hat{N}=\hat{c}^a\partial_aH(\hat{\varphi}), \qquad \hat{\bar{N}}=
\hat{\bar{c}}_a\omega^{ab}\partial_bH(\hat{\varphi})
\end{displaymath}
are further conserved charges \cite{quattro}. The expansion (\ref{29-2}) holds also if we replace 
the operators with the corresponding $c$-number variables, i.e.:
\be
\displaystyle H[\Phi^a]=H[\varphi^a]+\theta N+\bar{N}\bar{\theta}-i\bar{\theta}\theta
{\cal H}(\varphi^a,\lambda_a,c^a,\bar{c}_a). \label{29-3}
\ee
It is interesting to note that the first term in the expansion in $\theta,\bar{\theta}$
on the RHS of (\ref{29-3}) is $H(\varphi^a)$ which generates the dynamics in the standard 
{\it phase space} $\varphi^a$ while the last term is ${\cal H}$, which generates 
the dynamics in the extended phase space $(\varphi^a,\lambda_a,c^a,\bar{c}_a)$.

\section{Dequantization in the coordinate polarization and supertime}

In this section, which is the central one for our project, we will study 
the role of the superphase space variables $\Phi^a$ at the Lagrangian and path 
integral level.

We have seen that, for what concerns the Hamiltonians, relation (\ref{29-3}) holds:
\begin{displaymath}
H[\Phi]=H[\varphi]+\theta N+\bar{N}\bar{\theta}-i\bar{\theta}\theta {\cal H}
\end{displaymath}
which implies that
\be
\displaystyle i \int d\theta d\bar{\theta} H[\Phi]={\cal H}. \label{32-1}
\ee
An analog of this relation at the Lagrangian level does not hold exactly. The reason \cite{ann} is the presence in the Lagrangian of the kinetic terms
$p\dot{q}$, which are not present in $H$, i.e.:
\begin{displaymath}
\displaystyle L(p,q)=p\dot{q}-H(p,q). \label{32-2}
\end{displaymath}
Replacing $q$ and $p$ in the Lagrangian $L$ with the superphase space variables $\Phi^q$ and $\Phi^p$, the analog of Eq. (\ref{32-1}) becomes the following:
\be
\displaystyle i\int d\theta d\bar{\theta} L(\Phi)={\cal L}-
\frac{d}{dt}(\lambda_{p_i}p_i+i\bar{c}_{p_i}c^{p_i}), \label{33-1}
\ee
where ${\cal L}$ is the Lagrangian of the CPI given by Eq. (\ref{14-1a}),
and the $\lambda_{p_i}$, $\bar{c}_{p_i}$ and $c^{p_i}$ are the second half of the variables
$\lambda_a$, $\bar{c}_a$ and $c^a$. From now on we will change our notation for the superphase space variables: instead of writing 
\begin{displaymath}
\displaystyle \Phi^a\equiv \varphi^a+\theta c^a+\bar{\theta}\omega^{ab}\bar{c}_b+i\bar{\theta}\theta
\omega^{ab}\lambda_b
\end{displaymath}
we will explicitly indicate the $q$ and $p$ components in the following manner:
\be
\displaystyle \Phi^a=\begin{pmatrix} Q_i \cr P_i
\end{pmatrix}\equiv \begin{pmatrix} q_i \cr p_i \end{pmatrix}+\theta \begin{pmatrix}c^{q_i} \cr
c^{p_i} \end{pmatrix}+\bar{\theta}\begin{pmatrix} \bar{c}_{p_i} \cr -\bar{c}_{q_i}
\end{pmatrix}+i\bar{\theta}\theta \begin{pmatrix} \lambda_{p_i} \cr
-\lambda_{q_i}\end{pmatrix}, \label{33-2}
\ee
where $i=(1,\cdots, n)$, and $a=(1,\cdots, 2n)$. 

Going now back to (\ref{33-1}) we could write it as follows: 
\be
\displaystyle {\cal L}=i\int d\theta d\bar{\theta} L[\Phi]+\frac{d}{dt}(\lambda_pp
+i\bar{c}_pc^p), \label{33-3}
\ee 
where we have dropped the indices ``$i$'' appearing on the extended phase space variables. The expression of ${\cal L}$
which appears in (\ref{33-3}) can be used in (\ref{19-3}) and we get
\bea
\displaystyle \langle \varphi,c,t|\varphi_{\ssc 0},c_{\ssc 0},t_{\ssc 0}\rangle &=& 
\int {\mathscr D}^{\prime\prime}\varphi{\mathscr D}\lambda{\mathscr D}^{\prime\prime}c{\mathscr D}
\bar{c} \, \exp i\int_{t_0}^t d\tau{\cal L}= \nonumber\\
&=& \int {\mathscr D}^{\prime\prime}\varphi {\mathscr D}\lambda {\mathscr D}^{\prime\prime}c
{\mathscr D}\bar{c} \, \exp \Bigl[ i\int_{t_0}^t i d\tau d\theta d\bar{\theta}L[\Phi]+(\textrm{s.t.})\Bigr]
\label{34-1}
\eea
where (s.t.) indicates the surface terms, which come from the total derivative appearing on the RHS of (\ref{33-3}) and has the form
\be
\displaystyle (\textrm{s.t.})=i\lambda_pp-i\lambda_{p_{\ssc 0}}p_{\ssc 0}-
\bar{c}_pc^p+\bar{c}_{p_{\ssc 0}}c^{p_{\ssc 0}}. \label{34-2}
\ee 
We indicate with $p_{\ssc 0}$ the $n$ components of the initial momenta.
The surface terms present in (\ref{34-1}) somehow spoil the beauty of formula (\ref{34-1}) but we can get rid of them by changing the basis of our Hilbert space.
We already showed that, besides the basis $| \varphi,c\rangle=| q,p,c^q,c^p\rangle$, we could introduce the ``mixed'' basis defined in (\ref{15-2}) by the states: $|q,\lambda_p,
c^q,\bar{c}_p\rangle$. We can then pass from the transition amplitude
$\langle q,p,c^q,c^p,t|q_{\ssc 0},p_{\ssc 0},c^{q_{\ssc 0}},\bar{c}_{p_{\ssc 0}},t_{\ssc 0}\rangle$ of (\ref{34-1}) 
to the mixed one 
\begin{displaymath}
\langle q,\lambda_p,c^q,\bar{c}_p,t|q_{\ssc 0},\lambda_{p_{\ssc 0}},
c^{q_{\ssc 0}},\bar{c}_{p_{\ssc 0}}, t_{\ssc 0}\rangle,
\end{displaymath}
which are related to each other as follows:
\bea
&& \displaystyle
\langle q,\lambda_p,c^q,\bar{c}_p,t|q_{\ssc 0},\lambda_{p_{\ssc 0}},c^{q_{\ssc 0}}, 
\bar{c}_{p_{\ssc 0}},t_{\ssc 0}\rangle= \label{35-1} \\
&& \displaystyle =\int dp\,dp_{\ssc 0}\,dc^p\,dc^{p_{\ssc 0}}\, e^{-i\lambda_pp}e^{\bar{c}_pc^p}
\langle q,p,c^q,c^p,t|q_{\ssc 0},p_{\ssc 0},c^{q_{\ssc 0}},c^{p_{\ssc 0}},t_{\ssc 0}\rangle 
e^{i\lambda_{p_{\ssc 0}}p_{\ssc 0}}
e^{-\bar{c}_{p_{\ssc 0}}c^{p_{\ssc 0}}}. \nonumber
\eea 
Replacing on the RHS of this formula the kernel $\langle q,p,c^q,c^p,t|q_{\ssc 0},p_{\ssc 0}, c^{q_{\ssc 0}},c^{p_{\ssc 0}},t_{\ssc 0}\rangle$ with its path integral expression given in (\ref{34-1}), we get the very {\it neat} expression:
\be
\displaystyle \langle q,\lambda_p,c^q,\bar{c}_p,t|q_{\ssc 0},\lambda_{p_{\ssc 0}}, 
c^{q_{\ssc 0}},\bar{c}_{p_{\ssc 0}},t_{\ssc 0}\rangle=
\int {\mathscr D}^{\prime\prime}Q{\mathscr D}P\, \exp \Bigl[i\int_{t_0}^t id\tau d\theta d\bar{\theta}L(\Phi)\Bigr]
\label{35-2}
\ee
where 
\be
\displaystyle {\mathscr D}^{\prime\prime}Q{\mathscr D}P\equiv {\mathscr D}^{\prime\prime}
q{\mathscr D}p{\mathscr D}^{\prime\prime}\lambda_{p}{\mathscr D}\lambda_{q}
{\mathscr D}^{\prime\prime}c^{q}{\mathscr D}c^{p}{\mathscr D}^{\prime\prime}\bar{c}_{p}
{\mathscr D}\bar{c}_{q}. \label{35-3} 
\ee
We would like to point out three ``interesting'' aspects of Eq. (\ref{35-2}). 
\begin{enumerate}
\item[{\bf 1.}] Note that the surface terms of Eq. (\ref{34-1}) have disappeared in Eq. (\ref{35-2});
\item[{\bf 2.}] The measure in the path integral (\ref{35-2}) is the same
as the measure of the {\it quantum path integral} \cite{uno}, which is
\be
\displaystyle \langle q,t|q_{\ssc 0},t_{\ssc 0}\rangle =\int {\mathscr D}^{\prime\prime}q{\mathscr D}p \;
\exp \frac{i}{\hbar} \int d\tau\, L[\varphi]. \label{36-1} 
\ee
What we mean is that both in (\ref{35-2}) and in (\ref{36-1}) the integration in $p$ and $P$ is done
even over the initial and final variables while the integration in $q$ and $Q$ is done only over the intermediate 
points between the initial and final ones. The reason for the notation (\ref{35-3}) should be clear from the 
fact that the superphase space variables $Q$, $P$ are defined as
\be
\begin{array}{l}
Q \equiv q+\theta c^{q} +\bar{\theta}\bar{c}_{p}+i\bar{\theta}\theta \lambda_{p} \medskip \\
P\equiv p+\theta c^{p}-\bar{\theta}\bar{c}_{q}-i\bar{\theta}\theta \lambda_{q},
\label{36-2}
\end{array}
\ee
so the integration over $Q$, $P$ means the integration over the elements $(q,c^{q},
\bar{c}_{p},\lambda_{p})$ and $(p,c^{p},
\bar{c}_{q},\lambda_{q})$ which make up $Q$ and $P$ respectively;
\item[{\bf 3.}] The function $L$, which enters both the QM path integral (\ref{36-1}) 
and the CM one (\ref{35-2}), is the same. The only difference is that in QM (\ref{36-1}) the variables entering $L$ are the normal phase space variables while in CM (\ref{35-2}) they are the superphase space variables $\Phi^a=(Q,P)$.
\end{enumerate}
Let us now proceed by noticing that formally we can rewrite (\ref{35-2}) as
\be
\displaystyle \langle Q,t|Q_{\ssc 0},t_{\ssc 0}\rangle \equiv \int {\mathscr D}^{\prime\prime}Q{\mathscr D}P\,
\exp \left[i\int_{t_0}^t i d\tau d\theta d\bar{\theta} L[\Phi]\right] \label{36-3}
\ee
where we have defined the ket $|Q\rangle$ as the common eigenstate of the operators
$\hat{q}$, $\hat{\lambda}_p$, $\hat{c}^q$, $\hat{\bar{c}}_p$:
\be
\begin{array}{l}
\hat{q}|Q\rangle =q|Q\rangle,  
\medskip \\
\hat{c}^q|Q\rangle =c^q|Q\rangle,  
\end{array}
\qquad \quad
\begin{array}{l}
\hat{\lambda}_p|Q\rangle =\lambda_p|Q\rangle, 
\medskip \\
\hat{\bar{c}}_p|Q\rangle =\bar{c}_p|Q\rangle. \label{37-1}
\end{array}
\ee
So $|Q\rangle$ can be ``identified'' with the state $|q,\lambda_p,c^q,\bar{c}_p\rangle$ which appears
in (\ref{35-2}). The reader may not like this notation because in (\ref{36-2})  $Q$ contains the Grassmann variables $\theta$ 
and $\bar{\theta}$ which do not appear at all in (\ref{37-1}). Actually, from (\ref{37-1}) we can derive that the state $|Q\rangle$ is also an eigenstate of the supervariable $\hat{Q}$ obtained by turning the expression (\ref{36-2}) into an operator, i.e.:
\be
\displaystyle \hat{Q}(\theta,\bar{\theta})|Q\rangle =Q(\theta,\bar{\theta})|Q\rangle. \label{37-2}
\ee
This relation is just a simple consequence of (\ref{37-1}) as can be proved by expanding in $\theta$ and $\bar{\theta}$ both $\hat{Q}$ and $Q$ in (\ref{37-2}). One immediately sees that 
the variables $\theta$ and $\bar{\theta}$ make their appearance not in the state $|Q\rangle$ but in its 
eigenvalue $Q$ and in the operator $\hat{Q}$. We are now ready to compare (\ref{36-1}) and (\ref{36-3}). One is basically the central element of QM:
\be
\displaystyle \langle q,t|q_{\ssc 0}, t_{\ssc 0}\rangle =\int {\mathscr D}^{\prime\prime} q{\mathscr D}p
\exp \left[\frac{i}{\hbar} \int_{t_0}^t d\tau L[\varphi]\right], \label{38-1}
\ee 
while the other is the central element of CM (formulated \`a la KvN or \`a la CPI):
\be
\displaystyle \langle Q,t|Q_{\ssc 0},t_{\ssc 0}\rangle =\int {\mathscr D}^{\prime\prime}Q{\mathscr D}P
\exp \left[ i\int_{t_0}^t i d\tau d\theta d\bar{\theta} \,L[\Phi]\right]. \label{38-2}
\ee
By just looking at (\ref{38-1}) and (\ref{38-2}), it is now easy to give some simple {\it rules}, which turn the {\it quantum} transition amplitude (\ref{38-1}) into the {\it classical} one (\ref{38-2}). The rules are:
\begin{enumerate}
\item[{\bf 1)}] {\it Replace in the QM case the phase space variables} $(q,p)$ {\it everywhere with the
superphase space ones} $(Q,P)$;
\item[{\bf 2)}] {\it Extend the time integration to the supertime integration multiplied by} $\hbar$
\be
\displaystyle \int d\tau \, \longrightarrow \, i\hbar \int d\tau d\theta d\bar{\theta}. \label{39-1}
\ee
\end{enumerate}
The reason for the appearance of the ``$i$'' on the RHS of (\ref{39-1}) is related to the complex nature of the Grassmann variables $\theta$ and $\bar{\theta}$, as explained in Ref. \cite{ann}. The reason for the appearance of $\hbar$ instead is related to the fact that in (\ref{38-2}), which is CM, there is no $\hbar$ and so in (\ref{39-1}) we need to introduce an $\hbar$ in order to cancel the one of Eq. (\ref{38-1}). From the dimensional point of view formula (\ref{39-1}) 
is correct because, as shown in Ref. \cite{ann}, the dimensions of $d\theta d\bar{\theta}$ are just the inverse of an action canceling in this manner the dimension of $\hbar$ appearing in front of the RHS in (\ref{39-1}).
This implies that both the LHS and the RHS of (\ref{39-1}) have the dimensions of a time. We will call the rules {\bf 1)} and {\bf 2)} above as dequantization rules.

Note that these {\it dequantization} rules are {\it not} the semiclassical or WKB limit of QM.
In fact we are not sending $\hbar \to 0$ in (\ref{38-1}) and what we get is not QM in the 
leading order in $\hbar$, like in the WKB method, but exactly CM in the KvN or CPI formulation.
We named this procedure ``dequantization'' because it is the inverse of ``quantization'' in the sense
that, while quantization is a set of rules which turn CM into QM, our rules {\bf 1)} and {\bf 2)}
turn QM into CM.
We called `` {\it geometrical}'' our procedure because it basically consists of a geometrical 
{\it extension} of both the base space given by time $t$, into the supertime $(t,\theta,\bar{\theta})$ 
and of the target space, which is phase space $(q,p)$ in QM, into a superphase space $(Q,P)$ in CM.
We used the expressions ``{\it base space}'' and ``{\it target space}'', as it is done nowadays
in strings and higher dimensional theories, where procedures of dimensional extension or dimensional
contraction are very often encountered. In those theories the procedures of dimensional extension are introduced to give a geometrical basis to the many extra fields present in grand-unified theories,
while the procedure of dimensional contraction is needed to come back to our four-dimensional world.
We find it amazing and thought-provoking that even the procedure of quantization (or dequantization)
can be achieved via a dimensional contraction (or extension).

In previous papers \cite{cinque} we have given brief presentations of these ideas but there we explored the inverse route, that is the one of {\it quantization}, which is basically how to pass from (\ref{38-2}) to (\ref{38-1}). This goal is achieved by a sort of dimensional reduction from the supertime $(t,\theta,\bar{\theta})$ to the time $t$, and from the superphase space $(Q,P)$ to the phase space $(q,p)$. In those papers \cite{cinque}
we thought of implementing the supertime reduction by inserting a 
$\delta(\bar{\theta})\delta(\theta)/\hbar$ into the weight appearing in (\ref{38-2}) but we found this method
a little bit awkward and that is why here we have preferred to explore the opposite 
route that is the one of {\it dequantization} which is brought about by a dimensional extension. Even if awkward to implement, the quantization route from (\ref{38-2}) to (\ref{38-1}) can be
compared with a well-known method of quantization known in the literature as {\it geometric quantization}
(GQ) \cite{sei}. We will not review it here but suffice it to say that it starts from the so-called  ``prequantization space'' (which is our space of KvN states $\psi(q,p)$), and from the {\it Lie derivative
of the Hamiltonian flow} (which is our $\hat{{\cal H}}$ of equation (\ref{16-3})), and, through a long set of steps, it builds up the Schr\"odinger operator and the Hilbert space of QM. Basically, in going from (\ref{38-2}) to (\ref{38-1}), we do the same because we go from the weight
$\exp i \int i dt d\theta d\bar{\theta} L[\Phi]$, which is the evolution via the Lie derivative 
operator $\exp -i\hat{{\cal H}} t$, to the weight $\exp \frac{i}{\hbar}\int
L[\varphi]$, which reproduces the evolution via the Schr\"odinger operator 
$ \exp -i\frac{\hat{H}}{\hbar}t$. Regarding the states we go from the KvN states 
$|Q\rangle $ to the Schr\"odinger ones $| q\rangle$ by just sending $\theta,\bar{\theta} \to 0$ in Eq. (\ref{37-2}). The difference with respect to GQ is that our KvN states contain
also the Grassmann variables $c$ and $\bar{c}$, which are not contained in the prequantization states of GQ. In GQ the reduction of the KvN states to the quantum ones is achieved via a procedure called
``polarization'' while the transformation of the Lie derivative into the quantum Schr\"odinger
operator is achieved via a totally different procedure, see Ref. \cite{sei} for details. In our opinion it is somehow unpleasant that in GQ states and operators are ``quantized'' via two totally different procedures. This is not so anymore in our functional approach, which brings (\ref{38-2}) into (\ref{38-1}). It is in fact the dimensional reduction both in $t$ 
\be
\displaystyle i\hbar \int dt d\theta d\bar{\theta} \, \longrightarrow \, \int dt \label{43-1}
\ee 
and in phase space
\be
\displaystyle (Q,P) \, \longrightarrow \, (q,p), \label{43-2}
\ee
which at the same time produces the right operators (from the classical Lie derivative to the Schr\"odinger operator)
and the right states (from the KvN states $|Q\rangle$ to the quantum ones $| q\rangle$). So we do not need two
different procedures for operators and states but just a single one. Actually the dimensional reduction contained in (\ref{43-1}) and (\ref{43-2}) can be combined in a single operation, that is the one of {\it shrinking to zero} the variables $\theta,\bar{\theta}$, i.e.: $(\theta,\bar{\theta})\,\to \, 0$.
This not only brings the integration $\int dt d\theta d\bar{\theta}$ to $\int dt$ but, remembering 
the form of $(Q,P)$ i.e. (\ref{36-2}), it also brings 
\begin{displaymath}
\displaystyle Q\, \longrightarrow \, q.
\end{displaymath} 
Because of this, it reduces the KvN states $| Q\rangle$ to the quantum ones $| q\rangle$, which are a basis for the quantum Hilbert space in the Schr\"odinger representation. 
Note the difference with the GQ procedure: there one starts with the states $| q,p\rangle$ and the ``$p$'' is removed through a long set of steps, known as polarization \cite{sei}. In our approach instead we first replace the ``$p$'' in the $| q,p \rangle$ states with the $\lambda_p$ via the {\it Fourier transform} presented in (\ref{35-1}) and then remove the $\lambda_p$ by {\it sending} $(\theta,\bar{\theta})\to 0$. We want to stress again that the same two steps, {\bf 1)} {\it Fourier transform} and {\bf 2)} {\it sending} $\theta,\bar{\theta} \to 0$, which polarizes the states, are the same ones which turn the {\it classical } evolution into the {\it quantum} one. In fact step {\bf 1)} takes away the surface terms in (\ref{34-1}) bringing the weight to be of the same form as the quantum one and step {\bf 2)}, sending $(\theta,\bar{\theta})\to 0$, brings the classical weight into the quantum one. We feel that this coincidence of the two procedures, for the states and the operators, was not noticed in GQ because there they did not use the partners of time $(\theta,\bar{\theta})$ and the functional approach. This coincidence is quite interesting because (besides a trivial Fourier transform) it boils down to be a {\it geometrical} operation: the dimensional reduction from supertime $(t,\theta,\bar{\theta})$ to time $t$. We feel that this is really the {\it geometry} at the heart of geometric quantization and of quantum mechanics in general. 
 
Before concluding this section, we would like to draw again the reader's attention to three crucial things, which made the whole procedure work as nicely as it did. The {\it first} one is that the classical weight $\int dt {\cal L}$ and the quantum one $\int dt L$ belong, modulo surface terms, to the same multiplet. In fact if we expand $S[\Phi]= \int dt L[\Phi]$ in $\theta$ and $\bar{\theta}$ we get:
\begin{equation}
\displaystyle S[\Phi] = \int dt L(\varphi) + \theta {\cal T}(\varphi,\lambda,c,\bar{c})
+\bar{\theta} {\cal V} (\varphi, \lambda,c,\bar{c}) +i\theta \bar{\theta} \left( \int dt {\cal L} (\varphi,\lambda,c,\bar{c}) +\textrm{s.t.} \right), \label{51-1} 
\end{equation}
 where the functions ${\cal T}$ and ${\cal V}$ are the analog of the $N$ and $\bar{N}$ which appeared in (\ref{29-2}). It is not important to write down the explicit form of ${\cal T}$ and ${\cal V}$ but to note that in (\ref{51-1}) the weights entering respectively the QPI, i.e. $\int dt L(\varphi)$, and the CPI, i.e. $\int dt {\cal L}(\varphi, \lambda, c,\bar{c})$, belong to the same multiplet (modulo surface terms). Somehow we can say that ``{\it God has put CM and QM in the same multiplet (modulo surface terms)}''. 
 
The {\it second thing} to note is that in the statement in italics and between quotation marks written above we can even drop the ``modulo surface terms'' part of the sentence. In fact those surface terms are crucial because, combined with the extra surface terms coming from the partial Fourier transforms of Eq. (\ref{35-1}), they give exactly the classical weight that goes into the quantum one by the single procedure of sending $\theta, \bar{\theta} \to 0$. 
 
The {\it third crucial thing} to note is that the partial Fourier transform of Eq. (\ref{35-1}), which produces the extra surface terms exactly needed to implement the procedure above, it produces also the right classical states that go into the quantum ones by the same process of sending $\theta, \bar{\theta} \to 0$. This set of ``incredible coincidences'' works both for the coordinate and the momentum representations and also for the coherent states ones (see Ref. \cite{ann} for details). 

\section{Conclusions and Outlooks}

We will not summarize here what we did in this talk. We will only write down the dequantization rules, which are now in their complete form, and next we will outline some physical ideas that we are pursuing at the moment. 

We can summarize our dequantization procedure as follows. In order to pass from QM to CM one must apply the following rules:
\begin{itemize}
\item[{\bf 1)}] Given a QM object, in case it is not already written in functional form, build from it a {\it functional} expression, which either gives the action of that object on the states, or from which the object itself can be derived;
\item[{\bf 2)}] Next, in that functional expression perform the following replacements:
\begin{itemize}
\item[{\bf A)}] the time integration with a proper supertime integration:
\begin{displaymath}
\displaystyle \int dt \; \longrightarrow \; i\hbar \int dt d\theta d\bar{\theta},
\end{displaymath}
\item[{\bf B)}] the phase space variables with the superphase space variables:
\begin{displaymath}
\displaystyle \varphi^a(t) \; \longrightarrow \; \Phi^a(t,\theta,\bar{\theta}).
\end{displaymath}
\end{itemize}
\end{itemize}

Another possibility to connect classical and quantum mechanics in a ``{\it geometrical}'' manner is the following one. We can build a generalized path integral in the enlarged space invariant under the super-diffeomorphisms in $(t,\theta,\bar{\theta})$ by introducing a metric $g$ or, equivalently, a vierbein $E$ in the space $(t,\theta, \bar{\theta})$. 
It is then possible to show that both the CPI and the QPI can be reproduced by choosing two different sets of metrics. 

Since we want to implement a new path integral, invariant under diffeomorphisms in the supertime, we are very close to the spirit of the first papers on supergravity \cite{arnnat}. Consequently, we can use some of their structures and results. 
First of all, let us choose the analog of the Minkowski metric in our ``space-time'' which is made up by one even variable $t$ and two odd variables $\theta$ and $\bar{\theta}$. Such a metric must be symmetric in the even-even part and antisymmetric in the odd-odd one. We can choose for example:
\begin{displaymath}
\displaystyle \eta_{\scriptscriptstyle{AB}}=\begin{pmatrix}
1 & 0 & 0 \cr 0 & 0 &-1 \cr 0 & 1 & 0
\end{pmatrix}.
\end{displaymath}
The analogue of the Lorentz group is given by the orthosymplectic group OSp(1,2) whose generators are:
\begin{equation}
\begin{array}{l}
\displaystyle X_{\scriptscriptstyle 1}=-\bar{\theta}\partial_{\theta}, \qquad 
X_{\scriptscriptstyle 2}=\theta \partial_{\bar{\theta}}, \qquad X_{\scriptscriptstyle 3}=-\frac{1}{2}(\bar{\theta}\partial_{\bar{\theta}}-\theta\partial_{\theta}) \\
\displaystyle X_{\scriptscriptstyle 4}=-\frac{1}{\sqrt{2}}(\bar{\theta}\partial_t-t\partial_{\theta}), \qquad X_{\scriptscriptstyle 5}=\frac{1}{\sqrt{2}}(\theta\partial_t+t\partial_{\bar{\theta}}) \label{osp12}.
\end{array}
\end{equation}
The operators $X_i$ of Eq. (\ref{osp12}) leave invariant the distance of a point of the supertime $z^{\scriptscriptstyle A}\equiv (t,\theta,\bar{\theta})$ from the origin, i.e. the following quantity: $
\displaystyle F\equiv z^{\scriptscriptstyle A} \eta_{\scriptscriptstyle AB}z^{\scriptscriptstyle B}=t^2-2\bar{\theta}\theta.$
Once we have defined the Minkowski metric, the relationship between the metric $g$ and vierbein $E$ is given by the following equation: 
\begin{equation}
g_{\scriptscriptstyle MN}=E^{\scriptscriptstyle A}_{\scriptscriptstyle M}\eta_{\scriptscriptstyle AB} (-1)^{\scriptscriptstyle (1+B)N}E^{\scriptscriptstyle B}_{\scriptscriptstyle N}.  \label{metvier}
\end{equation}

Before going on, let us notice that in Eqs. (\ref{38-1}) and (\ref{38-2}) we can easily integrate away the conjugate momenta $p$ and $P$ respectively. The results are the QPI and the CPI expressed entirely in terms of the configurational variables $q$ and $Q$
\begin{eqnarray}
&& \displaystyle \langle q,t|q_{\ssc 0}, t_{\ssc 0}\rangle =\int {\mathscr D}^{\prime\prime} q\, \exp \left[ \frac{i}{\hbar}\int_{t_0}
^t d\tau 
\left(\frac{1}{2}\partial_{\tau} q \partial_{\tau} q -V(q) \right) \right] \medskip \label{conf} \\
&& \displaystyle \langle Q, t| Q_{\ssc 0}, t_{\ssc 0}\rangle =\int {\mathscr D}^{\prime\prime}Q \, \exp \left[ i \int_{t_0}^t i d\tau d\theta d\bar{\theta}\, \left(\frac{1}{2}\partial_{\tau} Q(\tau,\theta,\bar{\theta})\partial_{\tau}Q(\tau,\theta, \bar{\theta})-V(Q)\right)\right].
\nonumber
\end{eqnarray}
The weights of the QPI and the CPI of Eq. (\ref{conf}) are different. Nevertheless, as we are going to prove, they can be considered as particular cases of the following
action, invariant under generic reparametrizations of the supertime $(\tau,\theta, \bar{\theta})$ \cite{brink}:
\begin{equation}
\displaystyle S=i\int \textrm{d}\tau\textrm{d}\theta \textrm{d}\bar{\theta} \, E \left[ \frac{1}{2}D_{\tau}Q(\tau,\theta,\bar{\theta})D_{\tau}Q({\tau},\theta,\bar{\theta})-V(Q)\right], \label{action}
\end{equation}
where $D_{\tau}\equiv E^{\scriptscriptstyle{M}}_{\tau}\partial_{\scriptscriptstyle{M}}$ and $E\equiv\textrm{sdet}(E^{\scriptscriptstyle{A}}_{\scriptscriptstyle{M}})$ are functions of the inverse matrix vierbeins $E^{\scriptscriptstyle{M}}_{\scriptscriptstyle{A}}$ and $E^{\scriptscriptstyle{A}}_{\scriptscriptstyle{M}}$, moreover the superfield $Q(\tau,\theta,\bar{\theta})$ transforms as a scalar under superdiffeomorphisms. 
From Eq. (\ref{action}) we see that, in order to reproduce the potential term $V(Q)$ of the CPI in the second line of (\ref{conf}), it is crucial to impose the condition $E=1$. Moreover, the right kinetic terms of the CPI are reproduced if the following equation holds:
\begin{equation}
\displaystyle D_{\tau}QD_{\tau}Q=\partial_{\tau}Q\partial_{\tau}Q. \label{kinterm}
\end{equation}
Let us parametrize the vierbein matrix $E^{\scriptscriptstyle M}_{\scriptscriptstyle A}$ as follows:
\begin{equation}
\displaystyle E^{\scriptscriptstyle{M}}_{\scriptscriptstyle{A}}\equiv 
\begin{pmatrix}
a & \alpha & \beta \cr
\gamma & b & c \cr
\delta & d & e  \label{vier}
\end{pmatrix}
\end{equation}
where $a, b, c, d$ and $e$ are Grassmannian even and $\alpha, \beta, \gamma$ and $\delta$ are Grassmannian odd. The condition (\ref{kinterm}) is satisfied iff $\alpha=\beta=0$ and $a=\pm 1$. So only the second and the third line of the vierbein remain free:
\begin{equation}
\displaystyle E^{\scriptscriptstyle{M}}_{\scriptscriptstyle{A}}\equiv 
\begin{pmatrix}
\pm 1 & 0 & 0 \cr
\gamma & b & c \cr
\delta & d & e. \label{matCPI}
\end{pmatrix}
\end{equation}
Let us remember that the vierbein above must satisfy also the condition on the determinant, i.e. $\textrm{sdet}\, E^{\scriptscriptstyle{M}}_{\scriptscriptstyle{A}}=1$. This implies:
\begin{equation}
\displaystyle \textrm{det} \begin{pmatrix}
 b & c \cr
 d & e 
\end{pmatrix}=\pm 1 \, \Rightarrow \, be-cd =\pm 1. \label{det}
\end{equation}
If we rewrite $b, c, d, e$ in terms of their ``bodies'' $b_{\scriptscriptstyle B}$ and souls
\cite{dieci} $b_{\scriptscriptstyle S}$, e.g. $b=b_{\scriptscriptstyle{B}}+b_{\scriptscriptstyle{S}}\bar{\theta}\theta$,
then Eq. (\ref{det}) has as solutions:
\begin{equation}
\displaystyle b_{\scriptscriptstyle{B}}=\frac{\pm 1+c_{\scriptscriptstyle{B}}d_{\scriptscriptstyle{B}}}{e_{\scriptscriptstyle{B}}}, \qquad 
b_{\scriptscriptstyle{S}}=\frac{\mp e_{\scriptscriptstyle{S}}-c_{\scriptscriptstyle{B}}d_{\scriptscriptstyle{B}}e_{\scriptscriptstyle{S}}+c_{\scriptscriptstyle{B}}d_{\scriptscriptstyle{S}}e_{\scriptscriptstyle{B}}+c_{\scriptscriptstyle{S}}d_{\scriptscriptstyle{B}}e_{\scriptscriptstyle{B}}}{e_{\scriptscriptstyle{B}}^2}, \qquad  e_{\scriptscriptstyle{B}}\neq 0
\label{one}
\end{equation}
or
\begin{equation}
\displaystyle c_{\scriptscriptstyle{B}}=\mp\frac{1}{d_{\scriptscriptstyle{B}}},\qquad 
c_{\scriptscriptstyle{S}}=\frac{\pm d_{\scriptscriptstyle{S}}+b_{\scriptscriptstyle{B}}d_{\scriptscriptstyle{B}}e_{\scriptscriptstyle{S}}}{d_{\scriptscriptstyle{B}}^2},  \qquad e_{\scriptscriptstyle{B}}= 0. \label{two}
\end{equation}
These are the only constraints that the entries of the matrix $E^{\scriptscriptstyle{M}}_{\scriptscriptstyle{A}}$ of Eq. (\ref{matCPI}) must satisfy in order to get the weight of the CPI from the action (\ref{action}). 


Let us now go back to the action (\ref{action}) and let us see if different choices of the vierbein can reproduce also the weight of the QPI. In order to get a potential $V(q)$ we have to choose the following superdeterminant:
\begin{equation}
\displaystyle E=-\frac{i\bar{\theta}\theta}{\hbar}. \label{sdet2}
\end{equation}
This creates a technical problem. In fact a Grassmann number with zero body, like the one of Eq. (\ref{sdet2}), does not admit an inverse \cite{dieci}. Let us bypass this problem by introducing a small ``regularizing'' parameter $\epsilon$ and defining a ``regularized'' superdeterminant:
\begin{equation}
\displaystyle E=\epsilon -\frac{i\bar{\theta}\theta}{\hbar}. \label{sdet3}
\end{equation}
The expression (\ref{sdet2}) and, as we will see, the weight of the QPI, can be obtained in the limit $\epsilon \to 0$. The inverse of (\ref{sdet3}) has the following expression:
\begin{equation}
E^{-1}=\frac{1}{\epsilon}+\frac{i\bar{\theta}\theta}{\epsilon^2\hbar}. \label{detqu}
\end{equation}
This is the determinant of the vierbein $E^{\scriptscriptstyle M}_{\scriptscriptstyle A}$ which enters the definition of the kinetic terms of the action (\ref{action}). Such an action becomes:
\begin{equation}
\displaystyle S=i\epsilon 
\int \textrm{d}\tau \textrm{d}\theta \textrm{d}\bar{\theta}\, \left[\frac{1}{2}D_{\tau}QD_{\tau}Q-V(Q)\right]
+\frac{1}{\hbar}\int \textrm{d}\tau\textrm{d}\theta \textrm{d}\bar{\theta} \, \bar{\theta}\theta
\left[\frac{1}{2}D_{\tau}QD_{\tau}Q-V(Q)\right]. \label{quant}
\end{equation}
The first term goes to zero when $\epsilon \to 0$ while the second term, by performing the Grassmann integrations in $\theta, \bar{\theta}$, reduces to
$\displaystyle \frac{1}{\hbar} \int \textrm{d}\tau \, \left[ \frac{1}{2} a_{\scriptscriptstyle B}^2 \partial_{\tau} q \partial_{\tau} q-V(q) \right]$.
This is the weight of the QPI, provided $a_{\scriptscriptstyle B}=\pm 1$. Other two conditions can be derived from Eq. (\ref{detqu}). Let us rewrite $E^{-1}$ as:
\begin{equation}
E^{-1}=\left(\pm 1 +a_{\scriptscriptstyle S}\bar{\theta}\theta-p\bar{\theta}\theta\right)
(q+r\bar{\theta}\theta) = \pm q+(a_{\scriptscriptstyle S}q-pq\pm r)\bar{\theta}\theta,
\label{detqua}
\end{equation}
where we have defined $\displaystyle p\bar{\theta}\theta \equiv \begin{pmatrix} \alpha &\beta \end{pmatrix} \begin{pmatrix} b & c \cr d & e \end{pmatrix}^{-1}
\begin{pmatrix} \gamma \cr \delta \end{pmatrix},$ and 
$q+r\bar{\theta}\theta \equiv \textrm{det}^{-1}\begin{pmatrix} b & c \cr d & e \end{pmatrix}.$
By equating (\ref{detqu}) and (\ref{detqua}) we get the constraint equations that have to be satisfied by the vierbeins. These equations and their solutions are not very enlightening. Nevertheless, we can put $a_{\scriptscriptstyle S}=\alpha=\beta=0$, as in the classical solution that we have found before, see Eq. (\ref{matCPI}). In this particular case it turns out that $p=0$ and the solutions of the constraint equations can be written as:
\begin{equation}
\displaystyle b_{\scriptscriptstyle{B}}=\frac{\pm \epsilon+c_{\scriptscriptstyle{B}}d_{\scriptscriptstyle{B}}}{e_{\scriptscriptstyle{B}}}, \quad 
b_{\scriptscriptstyle{S}}=\frac{\mp \epsilon e_{\scriptscriptstyle{S}}-c_{\scriptscriptstyle{B}}d_{\scriptscriptstyle{B}}e_{\scriptscriptstyle{S}}+c_{\scriptscriptstyle{B}}d_{\scriptscriptstyle{S}}e_{\scriptscriptstyle{B}}+c_{\scriptscriptstyle{S}}d_{\scriptscriptstyle{B}}e_{\scriptscriptstyle{B}}\mp(1-\epsilon)\frac{i}{\hbar} e_{\scriptscriptstyle B}}{e_{\scriptscriptstyle{B}}^2}, \quad  e_{\scriptscriptstyle B}\neq 0 \label{oneone}
\end{equation}
or
\begin{equation}
\displaystyle c_{\scriptscriptstyle{B}}=\mp\frac{\epsilon}{d_{\scriptscriptstyle{B}}},\qquad 
c_{\scriptscriptstyle{S}}=\frac{\pm \epsilon d_{\scriptscriptstyle{S}}+b_{\scriptscriptstyle{B}}d_{\scriptscriptstyle{B}}e_{\scriptscriptstyle{S}}\pm (1-\epsilon)\frac{i}{\hbar}d_{\scriptscriptstyle B}}{d_{\scriptscriptstyle{B}}^2}, \qquad e_{\scriptscriptstyle B}= 0. 
\label{twotwo}
\end{equation}
The interesting feature of these solutions is that they 
reproduce the weight of the QPI in the limit $\epsilon \to 0$, and the weight of the CPI in the limit $\epsilon \to 1$. In fact, in such a limit Eqs. (\ref{oneone})-(\ref{twotwo}) reduce to (\ref{one})-(\ref{two}). 

What we would like to do next is to characterize in a better way the sets of metrics which allow us to reproduce the weights of the CPI and of the QPI from the action (\ref{action}). A first natural step in order to do this is to derive from the vierbeins the associated metrics. For example, using Eq. (\ref{metvier}), we can deduce that the freedom in the choice of a metric compatible with the CPI can be parametrized by just five variables, as it emerges from the following expression:
\begin{equation}
g_{\scriptscriptstyle MN}=\begin{pmatrix}
1 & \mp \pi_1 \theta \mp \pi_2\bar{\theta} & \mp \pi_3\theta \mp \pi_4\bar{\theta} \cr
\pm \pi_1\theta\pm \pi_2\bar{\theta} & 0 & \mp 1 \mp\pi_5 \bar{\theta}\theta \cr 
\pm \pi_3\theta \pm \pi_4\bar{\theta} & \pm 1 \pm \pi_5\bar{\theta}\theta & 0 \label{mm}
\end{pmatrix}
\end{equation}
where:
\begin{equation}
\left\{ \begin{array}{l}
\pi_1 \equiv \gamma_{\theta}e_{\scriptscriptstyle B}-\delta_{\theta}c_{\scriptscriptstyle B}, \quad 
\pi_2 \equiv \gamma_{\bar{\theta}}e_{\scriptscriptstyle B}-\delta_{\bar{\theta}}c_{\scriptscriptstyle B}, \quad 
\pi_3 \equiv \delta_{\theta}b_{\scriptscriptstyle B}-\gamma_{\theta}d_{\scriptscriptstyle B},  \medskip \\
\pi_4 \equiv \delta_{\bar{\theta}}b_{\scriptscriptstyle B}-\gamma_{\bar{\theta}}d_{\scriptscriptstyle B}, \qquad 
\pi_5 \equiv \gamma_{\bar{\theta}}\delta_{\theta}-\gamma_{\theta}\delta_{\bar{\theta}}.
\end{array} \right.  \label{pi}
\end{equation}
In the previous equation $\gamma_{\theta}, \gamma_{\bar{\theta}}$ and $\delta_{\theta},
\delta_{\bar{\theta}}$ are the components of the expansion of $\gamma$ and $\delta$, e.g.: $\gamma=\gamma_{\theta}\theta+\gamma_{\bar{\theta}}\bar{\theta}$.
From Eq. (\ref{mm}) we can calculate the values of the Ricci scalar, \cite{arnnat}, associated with the CPI. Unfortunately, due to the freedom in the $\pi_i$ of Eq. (\ref{pi}), every value of the body of the Ricci scalar seems to be compatible with the metrics of the CPI. This means that it is not the value of the Ricci scalar which distinguishes the ``classical'' metrics from the ``quantum'' ones. Nevertheless we feel that there should exist some geometrical or topological invariant which plays a crucial role in discriminating between classical and quantum mechanics. It is clear in fact that we cannot go from the metrics of the CPI to the ones of the QPI via a superdiffeomorphism because the metrics of the QPI are not invertible. Is this ``non-invertibility'', caused maybe by some ``singularity'' located somewhere, that gives origin to the quantum effects? Can this singularity be found and better understood via geometrical and topological tools? 

Another issue is the one related to the Einstein equations associated with our supermetrics: something analog to $G^{\mu \nu}=kT^{\mu \nu}$. Some ``matter'' content should produce as solutions the metrics of the QPI and some other matter content the one of the CPI. But the matter content in both cases is given by the same fields $Q(t,\theta,\bar{\theta})$. So there must be some ``hidden'' matter responsible for the different $T^{\mu \nu}$. Is this ``hidden'' matter related to the singularity mentioned above? We will come back and expand on these issues in Ref. \cite{geomdue}.

\section*{Acknowledgments}
We wish to thank M. Reuter and E. Spallucci for many useful conversations regarding especially the last part of this paper. This work has been supported by funds from MIUR (Italy), the University of Trieste and INFN.

\end{document}